\newcommand{\Ppp}{\ensuremath{\mathrm{pp}}\xspace}
\newcommand{\Pt}{\ensuremath{\mathrm{t}}\xspace}
\newcommand{\Pb}{\ensuremath{\mathrm{b}}\xspace}
\newcommand{\Pq}{\ensuremath{\mathrm{q}}\xspace}
\newcommand{\PW}{\ensuremath{\mathrm{W}}\xspace}
\newcommand{\PZ}{\ensuremath{\mathrm{Z}}\xspace}
\newcommand{\PH}{\ensuremath{\mathrm{H}}\xspace}
\newcommand{\PX}{\ensuremath{\mathrm{X}}\xspace}
\newcommand{\ttbar}{\ensuremath{\mathrm{t\bar{t}}}\xspace}
\newcommand{\bbbar}{\ensuremath{\mathrm{b\bar{b}}}\xspace}
\newcommand{\ttbb}{\ensuremath{\mathrm{ttbb}}\xspace}
\newcommand{\ttHbb}{\ensuremath{\mathrm{ttH(bb)}}\xspace}
\newcommand{\ttZbb}{\ensuremath{\mathrm{ttZ(bb)}}\xspace}
\newcommand{\LH}{\ensuremath{\mathcal{L}}\xspace}
\newcommand{\vx}{\ensuremath{\mathbf{x}}\xspace}
\newcommand{\vxDNN}{\ensuremath{\mathbf{x}^{\mathrm{DNN}}}\xspace}
\newcommand{\vxDNNred}{\ensuremath{\vxDNN_{\mathrm{red}}}\xspace}
\newcommand{\va}{\ensuremath{\mathbf{a}}\xspace}
\newcommand{\pt}{\ensuremath{p_{\mathrm{T}}}\xspace}
\newcommand{\abseta}{\ensuremath{\left|\eta\right|}\xspace}
\newcommand{\GeV}{\ensuremath{\,\mathrm{GeV}}\xspace}

\newcommand{\Pell}{\ensuremath{\mathrm{\ell}}\xspace}
\newcommand{\Pnu}{\ensuremath{\mathrm{\nu_{\ell}}}\xspace}
\newcommand{\NTP}{\ensuremath{N_{\mathrm{TP}}}\xspace}
\newcommand{\NTPeff}{\ensuremath{N^{\mathrm{eff}}_{\mathrm{TP}}}\xspace}
\newcommand{\GNNoneL}{\ensuremath{\mathrm{GNN}_{\nconv=1}}\xspace}
\newcommand{\GNNtwoL}{\ensuremath{\mathrm{GNN}_{\nconv=2}}\xspace}
\newcommand{\expGNN}{\ensuremath{\mathrm{GNN}^{\uparrow}}\xspace}
\newcommand{\DNNoneL}{\ensuremath{\mathrm{DNN}_{\nconv=1}}\xspace}
\newcommand{\DNNtwoL}{\ensuremath{\mathrm{DNN}_{\nconv=2}}\xspace}

\newcommand{\restrDNN}{\ensuremath{\mathrm{DNN}^{\downarrow}}\xspace}
\newcommand{\restrDNNeff}{\ensuremath{\mathrm{DNN}^{\downarrow}_{\rm{eff}}}\xspace}
\newcommand{\NTPGNN}{\ensuremath{\NTP(\GNNtwoL)}\xspace}
\newcommand{\NTPDNN}{\ensuremath{\NTP(\DNNtwoL)}\xspace}
\newcommand{\NTPeffDNN}{\ensuremath{\NTPeff(\DNNtwoL)}\xspace}
\newcommand{\nemb}{\ensuremath{n}\xspace}
\newcommand{\nfeat}{\ensuremath{n_{\mathrm{feat}}}\xspace}
\newcommand{\nconv}{\ensuremath{k}\xspace}
\newcommand{\Minv}{\ensuremath{m}\xspace}
\newcommand{\DR}{\ensuremath{\Delta R}\xspace}
\newcommand{\DRinv}{\ensuremath{\Delta R^{-1}}\xspace}
\newcommand{\muAUC}{\ensuremath{\mu_{\mathrm{AUC}}}\xspace}
\newcommand{\DmuAUC}{\ensuremath{\Delta\mu_{\mathrm{AUC}}}\xspace}
\newcommand{\G}{\ensuremath{\mathcal{G}}\xspace}
\newcommand{\Irel}{\ensuremath{I_\mathrm{rel}^{\Delta R}}\xspace}
\newcommand{\mutrain}{\ensuremath{\mu_{\mathrm{Train}}}\xspace}
\newcommand{\muR}{\ensuremath{\mu_{R}}\xspace}
\newcommand{\none}{\textsf{none}\xspace}
\newcommand{\one}{\textsf{one}\xspace}
\newcommand{\rnd}{\textsf{rnd}\xspace}
\newcommand{\zero}{\textsf{zero}\xspace}
\newcommand{\orcid}[1]{\href{https://orcid.org/#1}{\hspace*{0.1em}\raisebox{-0.45ex}{\includegraphics[width=1em]{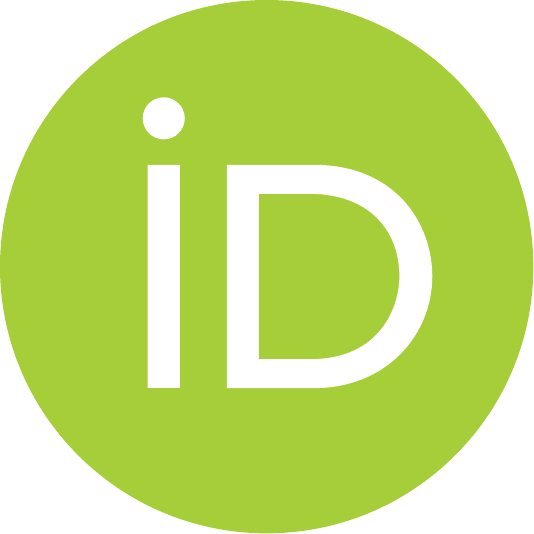}}}}

\RequirePackage{fix-cm}
\documentclass{svjour3}                     
\smartqed  
\usepackage{graphicx}
\usepackage{xspace}
\usepackage{amsmath}
\usepackage{amssymb}
\usepackage{todonotes}
\usepackage{multirow, bigdelim}
\usepackage{siunitx}
\usepackage{hyperref}
\hypersetup{
	colorlinks=true,
	linkcolor=black,
 	filecolor=black,
	urlcolor=black,
    citecolor=black,
}
\begin{document}

\title{A case study of sending graph neural networks back to the test bench for applications in high-energy particle physics
}

\titlerunning{Sending graph neural networks back to the test bench}        

\author{Emanuel Pfeffer\orcid{0009-0009-1748-974X}\and
        Michael Waßmer\orcid{0000-0002-0408-2811}\and
        Yee-Ying Cung \and
        Roger Wolf\orcid{0000-0001-9456-383X}\and
        Ulrich~Husemann\orcid{0000-0002-6198-8388}
}

\authorrunning{Emanuel Pfeffer \and Michael Waßmer \and Yee-Ying Cung \and Roger Wolf \and Ulrich Husemann} 

\institute{Emanuel Pfeffer\textsuperscript{1} (corresponding author) \at
           \email{emanuel.pfeffer@kit.edu}
           \and
           Michael Waßmer\textsuperscript{1} \at
           \email{michael.wassmer@kit.edu} 
           \and
           Yee-Ying Cung \at
           \email{yeeying.cung@web.de}
           \and
           Roger Wolf\textsuperscript{1} \at
           \email{roger.wolf@kit.edu}
           \and
           Ulrich Husemann\textsuperscript{1} \at
           \email{ulrich.husemann@kit.edu} 
           \and
           \textsuperscript{1} Karlsruhe Institute of Technology, Institute of Experimental Particle Physics, Karlsruhe, Germany\\
}

\date{ } 

\maketitle

\begin{abstract}
In high-energy particle collisions, the primary collision products usually decay further resulting in tree-like, hierarchical structures with a priori unknown multiplicity. At the stable-particle level all decay products of a collision form permutation invariant sets of final state objects. The analogy to mathematical graphs gives rise to the idea that graph neural networks (GNNs), which naturally resemble these properties, should be best-suited to address many tasks related to high-energy particle physics.
In this paper we describe a benchmark test of a typical GNN against neural networks of the well-established deep fully-connected feed-forward architecture. We aim at performing this comparison maximally unbiased in terms of nodes, hidden layers, or trainable parameters of the neural networks under study. 
As physics case we use the classification of the final state \PX produced in association with top quark-antiquark pairs in proton-proton collisions at the Large Hadron Collider at CERN, where \PX stands for a bottom quark-antiquark pair produced either non-resonantly or through the decay of an intermediately produced \PZ or Higgs boson.

\keywords{Graph Neural Networks \and Deep Neural Networks \and High-Energy Particle Physics \and LHC}
\end{abstract}

\section{Introduction}
\label{sec:introduction}

The continuous rise and flourish of deep learning has significantly impacted also the community of high-energy particle physics, where modern algorithms of deep learning ---mostly in the form of various neural network (NN) ar\-chi\-tec\-tures--- find applications as automation tools, for (multiclass) classification, parameter regression, or universal function approximation. The Large Hadron Collider (LHC) at CERN offers a unique test environment for such algorithms providing a large amount of independent identically distributed (i.i.d.) data from proton-proton (\Ppp) collisions under well controlled laboratory conditions. These data feature a rich hierarchical structure, optimally suited for the application of all kinds of general methods of statistical data analysis. Moreover, the underlying physics laws and statistical models, which have emerged over many decades of research, are scrutinized to a level that allows the reliable estimation of particle properties with a relative accuracy ranging far below the per-mille level, in rare cases even below $10^{-10}$~\cite{Muong-2:2023cdq}. This circumstance offers a toolbox for generating a large amount of perfectly known, complex, synthetic data, with a high relation to experimental observations, through the application of Monte Carlo (MC) methods~\cite{MC-method-1,MC-method-0}. These data are usually obtained as samples from an intractable though well-known likelihood function \LH. This setup provides a unique opportunity to thoroughly benchmark any kinds of machine learning (ML) algorithms under complex, real-life laboratory conditions.

At the LHC, data analysts strive for the application of more and more sophisticated ML-models with more and more not further processed ---and in this sense ``raw''--- input data. This strategy is fed by the belief that automated algorithms might find ways of extracting information of interest to the analyst, which are superior to selection strategies that are vulnerable to the bias of human prejudice. On the other hand, ML-algorithms should not be forced to learn already known and well-established physics principles, like symmetries inherent to the presented task. While such information can only be insufficiently passed through the necessarily finite training samples, it can be intrinsically incorporated either into the loss functions used for training, or in the NN architectures. 

At the large multi-purpose LHC experiments, ATLAS~\cite{Aad:2008zzm} and CMS~\cite{Chatrchyan:2008aa}, \Ppp collisions at a center-of-mass energy of, e.g., $13\,\mathrm{TeV}$ result in the creation of thousands of collision products to be recorded by the experiments. Primary collision products might decay further resulting in tree-like, hierarchical structures with a priori unknown multiplicity. The collision process can be described in a factorized approach: 

During the hard scattering process, the fundamental constituents of the protons, i.e., the quarks and gluons which are also collectively referred to as partons, interact via the fundamental interactions under investigation. We refer to the result of these interactions as the partonic final state. It cannot be observed directly in an experiment. Rather, each parton undergoes a series of theoretically well-known processes, setting in at lower energy scales, resulting in stable particles. The inverse problem usually subject of high-energy particle physics is to infer the presence and properties of the stable particles and eventually the partonic final state from their observable energy deposits in the detectors.

At the stable particle level all decay products of a collision form permutation invariant sets of final state objects, which may emerge from the collision in the form of collimated particle jets~\cite{Salam:2009jx}, forming well-suited proxies for strongly-interacting final state partons, or individual, spatially isolated particles, like leptons. From the preparation of the collision's initial state and energy and momentum conservation, physicists may infer the presence of non- or weakly-interacting particles, like neutrinos, in the collision's final state, through the principle of missing transverse energy (MET)~\cite{CMS:2019ctu}. A natural representation of this richly structured data is in the form of mathematical graphs \G, which are indeed also the basis of theoretical amplitude calculations of the quantum-mechanical wave function in the form of Feynman graphs~\cite{PhysRev.76.749}. 

 Within the high-energy particle physics community, this observation has lead to an increased interest in NNs based on mathematical graphs (GNNs)~\cite{GNN_original_1,GNN_original_2,GNN_original_3,GNN_original_4,GNN_original_5}, where nodes are usually identified by particles and edges potentially by relations across particles. A comprehensive review can be found in Ref.~\cite{Shlomi:2020gdn}. However, on closer inspection, several features of GNNs that count for seemingly obvious advantages reveal subtle challenges: 

\begin{itemize}
\item The mathematical model to match physics entities like stable particles to the graph \G of the GNN, which is subject to mathematical operations, bears ambiguities. The representation of particles by the nodes of \G appears as an obvious choice. This choice raises the question of the (potentially physical) meaning to the connecting edges, which could represent mother-daughter relations, or proximity in an arbitrarily defined space. This issue is emphasized once mathematical weights or even trainable parameters (TPs) used during node aggregation are assigned to the edges. 
\item Training and application imply the potentially complex and computationally time-consuming task of building \G, based on the physics inputs. 
\item In general, the more complex structure of the GNN compared to other NN architectures complicates the users' comprehension of how the GNN arrives at its prediction.  
\end{itemize}

In this paper we describe a comparison of typical GNN architectures with NN models based on the deep fully-connected feed-forward architecture (DNN), which has been studied intensely in the past. We aim at comparisons that are to best effort unbiased in terms of expressiveness and information provided to the NN models to solve a given task. In Section~\ref{sec:physics-task} we give an introduction to the task that serves as benchmark for this comparison. In Section~\ref{sec:nn-setup} the architectures and training setups of the NNs under study are described. In Section~\ref{sec:benchmark-tests} we present the results of the comparison. We conclude the paper with a summary in Section~\ref{sec:summary}.  

\section{Neural network task}
\label{sec:physics-task}

\subsection{Physics processes}

As benchmark for the comparison we use the classification of the final state~\PX produced in association with top (\Pt) quark-antiquark pairs (\ttbar) in \Ppp collisions at the LHC, where \PX stands for a bottom (\Pb) quark-antiquark pair (\bbbar) produced either through non-resonant gluon exchange or through the decay of a massive \PZ or Higgs (\PH) boson as intermediate particle, as discussed, e.g., in Ref.~\cite{CMS:2018hnq}. Under realistic conditions, the collision of interest might be overlaid by several tens of additional collisions, referred to as pileup. The complete detectable final state of a collision of interest, including pileup, is referred to as an event, whose feature vector \vx would be presented to the NN. An arbitrary number of such events may be generated synthetically by evaluating \LH of the full process via the MC method. In this study, we focus on the classification of the underlying hard process neglecting the effects of pileup.

\begin{figure}[b]
    \centering
    \includegraphics[width=0.32\textwidth]{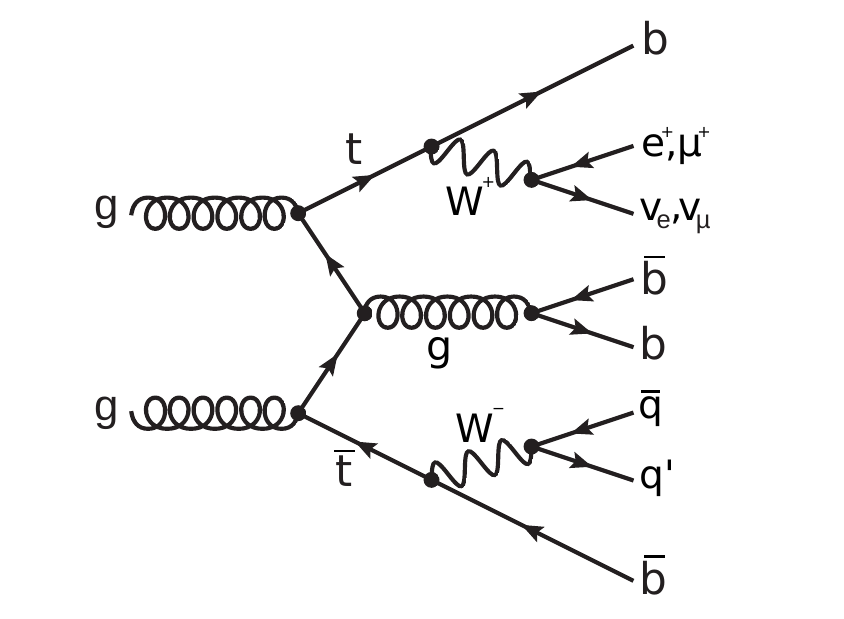}
    \includegraphics[width=0.32\textwidth]{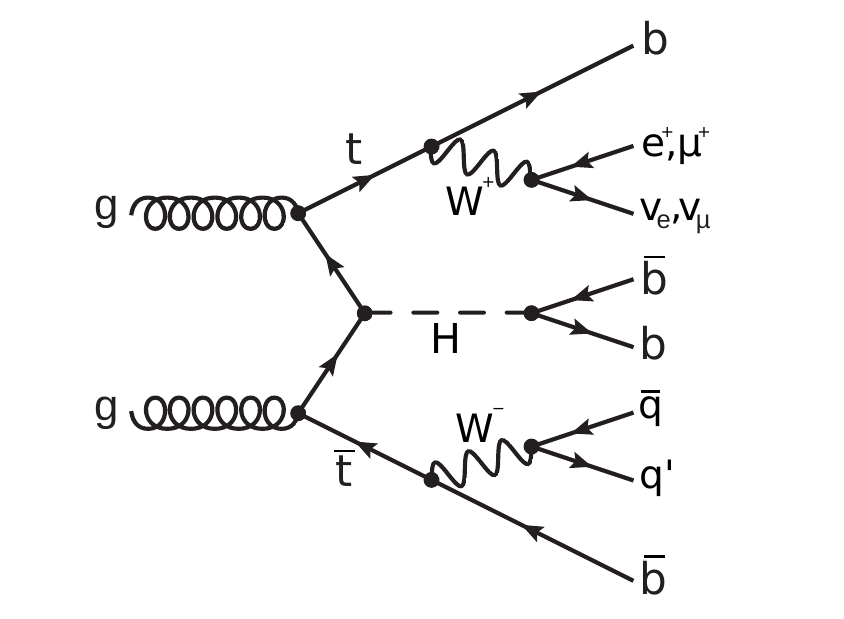}
    \includegraphics[width=0.32\textwidth]{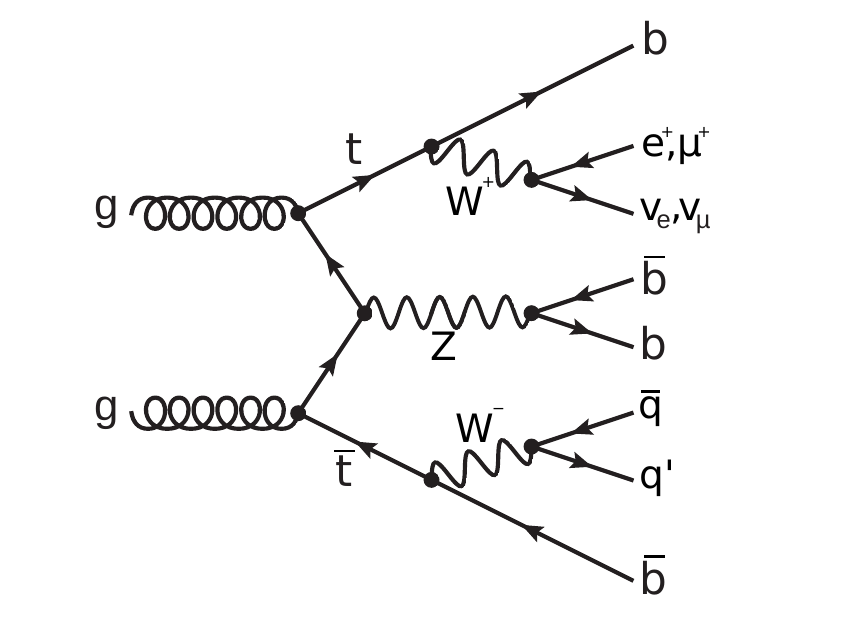}
    \caption{Exemplary Feynman diagrams for the processes of interest to this study: (left) \ttbb, (middle) \ttHbb, and (right) \ttZbb.}
    \label{fig:processes}
\end{figure}

The interest in the chosen classification task arose from studies of \PH production in association with \ttbar in the subsequent $\PH\to\Pb\bar{\Pb}$ decay (\ttHbb), for which \ttbar associated \PZ boson production ($\mathrm{ttZ}$) in the $\PZ\to\Pb\bar{\Pb}$ decay channel (\ttZbb) and non-resonant \bbbar production in association with \ttbar (\ttbb) are important background processes. Exemplary Feynman diagrams of these processes in leading-order (LO) of perturbation theory are shown in Figure~\ref{fig:processes}. 
The decay products, resembled by the outgoing lines in the diagrams, shown in Figure~\ref{fig:processes}, represent the partonic final state of interest to this study, which is the same for all processes. Therefore, the processes can only be distinguished by the kinematic properties of the particles, in particular \Pb quarks. This situation is complicated by the fact that the \Pt quark also decays into \Pb quarks radiating a quasi-real \PW boson with a branching fraction of nearly 100\%~\cite{Workman:2022ynf}. The \PW boson subsequently decays either into quarks, which further on form jets in the detector, or leptons. For the presented study the semi-leptonic \ttbar final state has been chosen, in which the \PW boson of one \Pt decays into an electron or muon, further on referred to as \Pell, and a corresponding neutrino \Pnu. The other \PW boson decays into quarks. Due to the radiation of additional gluons and the splitting of gluons into quark-antiquark pairs additional colored particles and consequently jets might emerge from the process. This constellation implies a richly structured final state of an event with at least four \Pb quark- and two predominantly light-quark-induced jets; an \Pell, which is spatially isolated from any other activity originating from the hard scattering process in the detector; and MET, due to the emitted \Pnu. The \Pb quark-induced jets, referred to as \Pb jets in the following, may be identified experimentally with a finite purity and efficiency, as, e.g., described in Ref.~\cite{CMS:2017wtu}; the methods of how to achieve this are not subject of this paper. 

\subsection{Sample preparation}
\label{sec:sample_preparation}

Samples for all processes in question have been generated synthetically from a corresponding likelihood \LH using the MC technique. The tools used for event generation are the matrix-element generator MadGraph5\_aMC@NLO~\cite{Alwall:2014hca,Alwall:2011uj} in version~2.9.9 interfaced with the Pythia event generator~\cite{Sjostrand:2014zea} in version~8.306 to map the partonic final state to the stable-particle level. All processes in consideration have been generated at LO in perturbative quantum chromodynamics (QCD), in the four-flavor scheme. The same setup has been used for the generation of all samples to avoid spurious differences due to the use of different generation tools. 

All generated events have been passed through a simplified simulation of the CMS detector as configured during the LHC Run-2 data-taking period in the years 2016--2018, using the DELPHES simulation package~\cite{deFavereau:2013fsa}. For this study only reconstructed leptons, jets, and MET are considered. All detected and reconstructed final-state objects have been selected to fulfill a set of selection criteria typically used for the analysis of data collected by the CMS experiment, as summarized in Table~\ref{tab:kin-selection}. These selection criteria comprise the following observables: 

\begin{itemize}
\item The transverse momentum \pt and pseudorapidity $\eta$ of the reconstructed \Pell and jets. 
\item A variable \Irel that corresponds to the scalar sum of energy deposits $E_{i}$ detected within the radius $\Delta R=\sqrt{\Delta\eta^2+\Delta\phi^2}$ around \Pell, divided by the \pt of \Pell, where $\Delta\eta$ refers to the difference between $E_{i}$ and \Pell in $\eta$ and $\Delta\phi$ to the corresponding difference in azimuthal angle $\phi$, based on the coordinate system deployed by CMS~\cite{Chatrchyan:2008aa}. For \Pell originating from $\PW\to\Pell\nu_{\Pell}$ low values of \Irel are expected.   
\item The output of a specific \Pb jet identification algorithm represented by the discrete observable $\beta\in\{\mathrm{0,1,2,3}\}$ indicating whether an object has been identified as a \Pb jet under a specific working point $\alpha$ ($\beta\geq\alpha$). The value of $\alpha$ represents ($\alpha=1$) loose, ($\alpha=2$) medium, and ($\alpha=3$) tight selection criteria, corresponding to a rate of non-\Pb jets wrongly identified as a \Pb jet (false-positive rate), of approximately 10\%, 1\%, and 0.1\%, respectively.
\end{itemize}

\begin{table}[t]
    \centering
    \caption{Selection requirements on the reconstructed final-state objects. The quantity \Irel corresponds to the scalar sum of energy deposits detected within the radius \DR around \Pell in $\eta$-$\phi$ divided by the \pt of \Pell, as defined in the text. A lower value of \Irel implies less activity in the spacial vicinity of \Pell, indicating that \Pell originates from $\PW\to\Pell\nu_{\Pell}$. The variable $\beta$ refers to the working points of a specific \Pb jet identification algorithm, as described in the text.}
    \begin{tabular}{lcccc}
       \hline
       Object  & $\pt\,(\GeV)$ & $\abseta$ & \Irel & $\beta$ \\
       \hline
       Electron & $\geq 25$ & $< 2.5$ & $< 0.12$ ($\DR=0.3$) & $-$ \\
       Muon & $\geq 25$ & $< 2.4$ & $< 0.25$ ($\DR=0.4$) & $-$ \\
       Jet (anti-$k_\text{T}$, $R_0$=0.4 \cite{Cacciari:2008gp}) & $\geq 20$ & $< 2.4$ & $-$ & $-$ \\
       \Pb jet & $\geq 20$ & $< 2.4$ & $-$ & $\beta\geq2$ \\
       \hline
    \end{tabular}
    \label{tab:kin-selection}
\end{table}
All events have been selected to exhibit at least six and not more than eight jets, at least four of which are assumed to be correctly identified as \Pb jets according to $\beta\geq2$, and exactly one \Pell, matching all selection criteria. The selection of one \Pell and six jets, of which at least four are identified as \Pb jets, is motivated by the partonic final state under study, as depicted in Figure~\ref{fig:processes}. The selection of up to two additional jets, increases the chance that the complete partonic final state can be matched to the selected jets. 

For each reconstructed jet the attempt is made to assign the initiating particle of the partonic final state, based on the distance $\DR$ between the jet and the corresponding parton. Only partonic final state objects with transverse momentum of $\pt>20\GeV$ and $\abseta<2.4$ are considered for this assignment. From the assignment five mutually exclusive jet classes are build, referring to the (ADDB) \Pb quarks not originating from any \Pt decay; the \Pb quark originating from the (HTB) hadronic and (LTB) leptonic \Pt decay, (HTQ) quarks originating from the hadronic \PW decay; and (NA) jets not assigned to the partonic final state. 

The assignment of the partonic final state to the reconstructed jets may be incomplete, for a given event. Events for which no or only one jet is assigned to the ADDB class are discarded from the training and test samples. In all other cases, if the assignment by \DR did not result in one jet of class LTB, one jet of class HTB, and two jets of class HTQ, the remaining not associated jets of class NA are ordered by decreasing \pt and the leading jets in \pt are re-assigned to these classes in the order of LTB, HTB, HTQ. A summary of all jet classes is given in Table~\ref{tab:jet-classes}

\begin{table}[t]
    \centering
    \caption{Requirements for the jet-class definition, according to the matching to the partonic final state, where $\mathrm{b_{add}}$ corresponds to a \Pb quark not originating from a \Pt decay, $\mathrm{b_{t_{had}}}$ ($\mathrm{b_{t_{lep}}}$) to a \Pb quark originating from $\Pt\to\Pb\PW(\Pq\Pq')$ ($\Pt\to\Pb\PW(\ell\nu_{\ell})$), and $\mathrm{q_{W_{had}}}$ to a quark originating from $\PW\to\Pq\Pq'$. We note that the classes ADDB and HTQ comprise two jets. Jets not assigned to any other class are assigned to the NA class and sorted by decreasing \pt. If no jet is found to fulfill the corresponding \DR criterion, the leading jet from the NA class is re-assigned to the classes HTB, HTL, HTQ, in that order.}
    \begin{tabular}{lll}
        \hline
        Class label & Assignment & Description \\
        \hline
        ADDB & $\DR(\mathrm{jet},\mathrm{b_{add}})<0.4$ & \Pb jets not from \Pt decays \\
        HTB & $\DR(\mathrm{jet},\mathrm{b_{t_{had}}})<0.4^{\dagger}$ & \Pb jet from $\mathrm{t_{had}}$ \\
        LTB & $\DR(\mathrm{jet},\mathrm{b_{t_{lep}}})<0.4^{\dagger}$ & \Pb jet from $\mathrm{t_{lep}}$ \\
        HTQ & $\DR(\mathrm{jet},\mathrm{q_{W_{had}}})<0.4^{\dagger}$ & \Pq jet from \PW \\
        NA & No match & Additional jet \\
        \hline
        \multicolumn{3}{l}{{\small $^{\dagger}$ If no assignment by \DR is found, the leading jet from NA is assigned.}}
    \end{tabular}
    \label{tab:jet-classes}
\end{table}

\subsection{Task definition}
\label{subsec:task_definition}

The NN models are supposed to perform a classification task, in which a given event should be assigned either to \ttbb, \ttHbb, or \ttZbb. The physics process from which the event was generated constitutes the ground truth information for this task. 

To simplify the task, the reconstructed jets are assumed to perfectly match the initiating partons, whenever possible, through the parton association algorithm, as described in Section~\ref{sec:sample_preparation}. The ADDB class contains jets which stem directly from the intermediate particle to distinguish the processes under study. The features of these jets are assumed to provide the most decisive contribution to the classification. 

\subsection{Training setup}
\label{subsec:training_setup}

All NN models are subject to a supervised training. The generated samples used for this are split into a training, validation, and test sample, containing 60, 20, and 20\% of the generated events, respectively. Event numbers, split by process and sample, are given in Table~\ref{tab:data_split}.

\begin{table}[b]
    \centering
    \caption{Numbers of events for each process, in the training, validation, and test samples.}
    \begin{tabular}{lcccc}
        \hline
        Process & Training & Validation & Test & Sum \\
        \hline
        \ttbb & 41650 & 13883 & 13803 & $\hphantom{0}69336$ \\
        \ttHbb & 99695 & 33329 & 33229 & 166253 \\
        \ttZbb & 88107 & 29272 & 29453 & 146832 \\
        \hline
    \end{tabular}
    \label{tab:data_split}
\end{table}

Each training is performed for an ensemble of ten statistically independent repetitions to obtain a rough estimate of the statistical spread of the trained models, due to random choices in the training setup. Performance measures of each model are reported as sample means $\mu$, of which the corresponding uncertainty $\Delta\mu$ is estimated from the square root of the sample variance. All repetitions are based on the same training, validation, and test samples. Due to the large number of events in these samples, randomization through data shuffling is assumed not to change the conclusions of the studies significantly.

Each training is performed on CPUs through a distributed computing infrastructure, where each training is performed on a dedicated CPU. We ensure that each step affected by random choices is based on a different random seed. The library used to build the GNNs is PyTorch Geometric v2.0.3~\cite{cite:PyTorch_Geometric} based on the PyTorch library~\cite{cite:PyTorch}. The same library is used, to construct the corresponding DNNs under study. Further parameter choices of the training setups are given in the upper part of Table~\ref{tab:hyperparameters}. They are the same for both NN architectures. 
\begin{table}[b]
    \centering
    \caption{Common parameter choices for the NN architectures under study and their training setup.}
    \begin{tabular}{ll}
        \hline
        Parameter & Setting(s) \\
        \hline
        Loss function & Binary cross-entropy \\
        Optimizer & Adam~\cite{Kingma:2014vow} ($\gamma=0.01$) \\
        Mini-batch size & 200 \\
        Maximum number of epochs & 200 \\
        Early-stopping & $\Delta\text{epochs}=15$, $\Delta\text{loss}=0.001$ \\
        \hline
        Use of weights and biases & Yes \\
        Number of outputs & 1 (binary) \\
        Activation function (for hidden layers) & ReLu \\
        Activation function (for output layer) & Sigmoid \\
        \hline
    \end{tabular}
    \label{tab:hyperparameters}
\end{table}

\section{Neural networks under study}
\label{sec:nn-setup}

\subsection{Architectures}

For the GNN models, a graph representation of the final state is obtained from the reconstructed jets, \Pell, and MET in an event. Each of these objects is represented by a node $i$ in \G. This set of nodes may be complemented by nodes for up to two more jets in the selection. Accordingly, \G has 8--10 nodes. For each~$i$, a vector~$\va_{i}$ of attributes with $\dim(\va_{i})=\nemb,\,\forall i$ is defined, forming the embedding space. At initialization time, the $\va_{i}$ are initialized by the feature vectors $\vx_{i}$ with $\dim(\vx_{i})=\nfeat,\,\forall i$. The following studies also comprise configurations with $\nemb>\nfeat$. In these cases, attributes in the $\va_{i}$ without correspondence in $\vx_{i}$ are set to zero (zero padding), at initialization time. 

\begin{figure}
    \centering
    \includegraphics[width=0.85\textwidth]{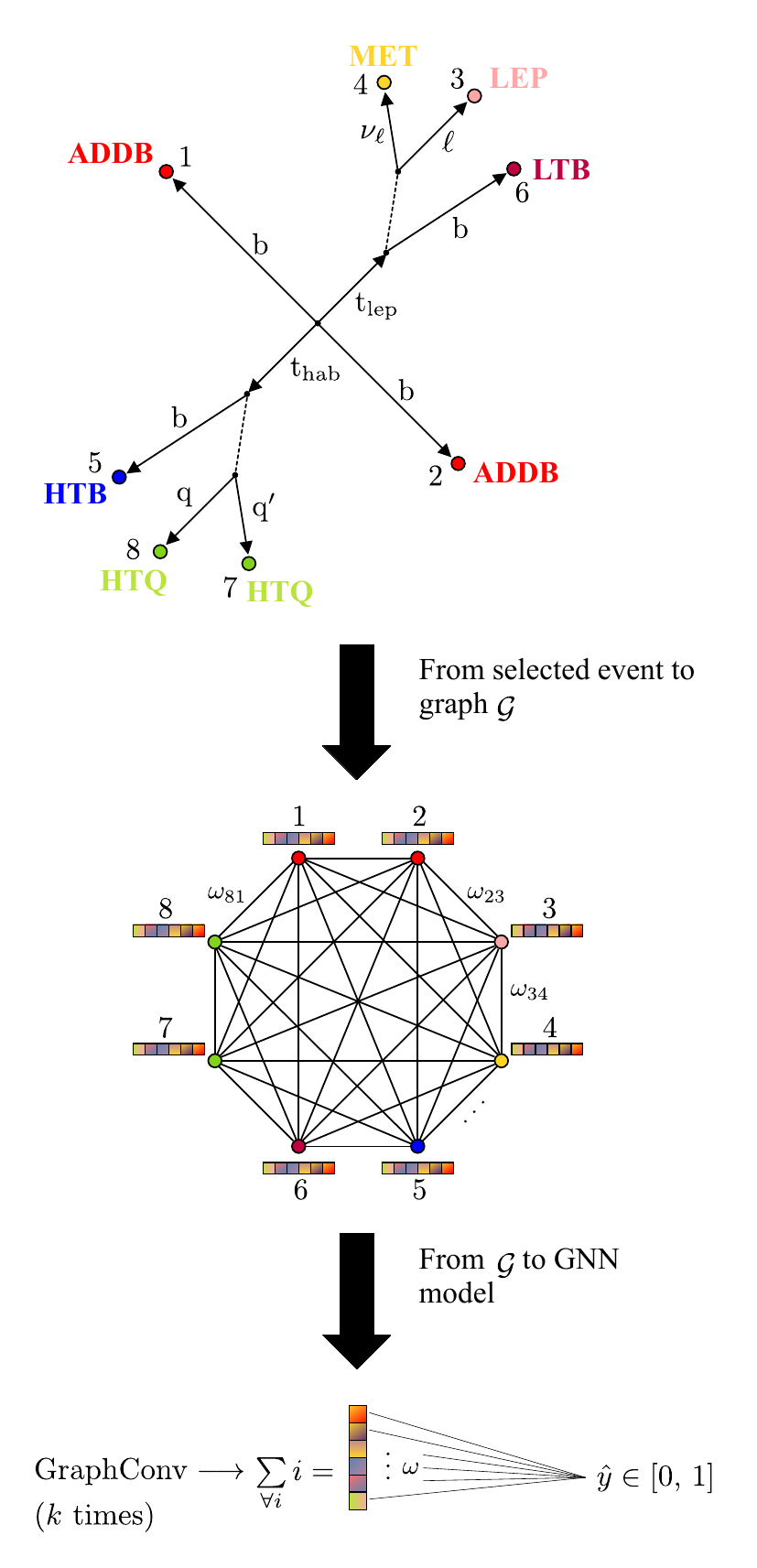}
    \caption{Translation of an (upper part) selected event (without jets in the NA jet-class in this case) into a (middle part) graph $\mathcal{G}$ and finally into the (lower part) GNN model. For the indication of the partonic final state we do not distinguish particles from anti-particles. The individual object-classes are indicated by different colors. The nodes of \G are labelled by $i=1\ldots 8$ and colored the same way as the object-classes. The boxes next to the nodes indicate the embedding space of the GNN model. The GNN output is indicated by $\hat{y}$.}
    \label{fig:GNN-model}
\end{figure}

All nodes are connected with edges, resulting in a fully-connected non-directed graph without self-loops. Relational information between nodes $i$ and $j$ may be assigned, in form of edge weights $\omega_{ij}$. For this purpose three physics motivated choices are made: (i) the invariant mass  \Minv; (ii) the distance \DR; and (iii) the reciprocal distance \DRinv of the connected final-state objects. In addition, the cases of a (\one) constant, (\rnd) random, and (\zero) no edge connection at all are studied, resulting in a total of six variants of edge connections. 
An illustration of a resulting graph for an event with eight nodes and no additional jets is given in Figure~\ref{fig:GNN-model}.

The GNN algorithm to process the graph data is based on the layered GraphConv operation, as introduced and described in Ref.~\cite{GraphConv}, using the sum over all $i$ as aggregation function. After initialization, \nconv GraphConv operations are applied, after which the resulting $\va_{i}$ are transformed into a single vector of length \nemb, averaging over all $i$. A linear combination of the components of this vector, which is scaled to values between 1 (indicating \ttbb as signal) and 0, for binary classification, eventually forms the output $\hat{y}$ of the GNN. A graphical illustration of this model is given in the lower part of Figure~\ref{fig:GNN-model}.

In a first study, the GNN are compared with corresponding DNN models with \nconv hidden layers, containing \nemb hidden nodes, each. The values of \nconv, \nemb, and the choice of weights, steering the exploitation of relational information by the GNN models are varied, resulting in 36 variants of parameter choices, as summarized in Table~\ref{tab:parameter_choices}. We note that here, as in the following, \nemb represents a tuple of length \nconv. For a GNN this tuple indicates the dimension of the embedding space per GraphConv operation; for a DNN it represents the number of nodes per hidden layer. 
\begin{table}[b]
    \centering
    \caption{Parameters varied for the comparison of GNN with corresponding DNN models, where \nemb corresponds to the dimension of the embedding space during a GraphConv operation (number of nodes in a hidden layer) and \nconv to the number of GraphConv operations (hidden layers) in the GNN (DNN) case. The choices of \nemb are motivated by the size of the input vector \nfeat to the GNN, as described in Section~\ref{sec:feature_space}.
    The choices of \one, \rnd, and \zero for the use of edge information in GNN models are compared to DNN models without relational information between individual objects. These DNN models are indicated by the label \none, in corresponding figures.}
    \begin{tabular}{ll}
        \hline
        Parameter & Setting(s) \\
        \hline
        \nconv & 1, 2 \\
        \nemb & 13, 26, 39 \\
        Edge weights ($\omega_{ij}$) & \Minv, \DR, \DRinv, \one, \rnd, \zero \\
        \hline
    \end{tabular}
    \label{tab:parameter_choices}
\end{table}
Other parameter choices related to the NN architectures under study or the setup of the NN training are made common and summarized in Table~\ref{tab:hyperparameters}. Special care is taken to compare the GNN with the corresponding DNN models on the same footing, especially in terms of information about the feature space presented to them, as discussed in the following section.  

\subsection{Presentation of the feature space}
\label{sec:feature_space}

Primary features passed to the NNs are the invariant mass ($m$), energy ($E$), $\eta$, and $\phi$ of each reconstructed final state object. The reconstructed final state objects comprise \Pell, MET, at least four \Pb jets, two additional jets, all of which are associated with the partonic final state, and potentially two more non-associated jets from the event selection. For MET the features $E$ and $m$ are set to zero. The primary features are complemented by $\beta$. The selection requires values of $\beta\geq2$ for identified \Pb jets. For \Pell and MET $\beta$ is set to zero. 

We note that both NN architectures may profit from additional information, which is not passed explicitly through $\vx_{i}$, but implicitly through the way the features are presented to the NNs. An obvious difference between the NN architectures arises from the fact that the GNN naturally supports processing of events with arbitrary object multiplicities. The $\vx_{i}$ are transferred to the GNN through the $\va_{i}$, during initialization. During the GNN processing the information per $i$ is aggregated over all nodes. In the DNN case such an aggregation step is absent. Instead, the $\vx_{i}$ are concatenated into an enlarged feature vector \vxDNN of length $10\times\nfeat$, comprising the $\vx_{i}$ of the eight reconstructed objects, of which all jets have been associated to the partonic final state, plus potentially two additional selected jets. The order in which these objects are concatenated has been chosen to follow the association to the partonic final state. For events that contain fewer than two jets in addition to those that have been matched to the partonic final state, the corresponding entries in \vxDNN are filled with zeros. We point out two subtleties, which are related to these choices: 

One subtlety, in favor of the GNN, arises from incorporating relational information through $\omega_{ij}$. This advantage is compensated for by appending equivalent information to \vxDNN. For up to ten selected objects in an event this results in up to 45 additional features. The choices of \one, \rnd, and \zero for the use of edge information in GNN models are compared to DNN models without relational information between individual objects. These DNN models are indicated by the label \none, in corresponding figures.

Another subtlety, related to the same fundamental difference, but this time in favor of the DNN, arises from the fact that through the concatenation of the $\vx_{i}$ into \vxDNN, according to the association to the partonic final state, the DNN receives extra information through the positions of the $\vx_{i}$ in \vxDNN which is not accessible to the GNN. This advantage is compensated for by adding information about the association of the $i$ to the partonic final state via one-hot encoding. For the five jet-classes defined in Table~\ref{tab:jet-classes} plus one label (LEP) for \Pell and one label (MET) for MET this extension increases \nfeat by seven, leading to the dimension of $\vx_{i},\,\forall i$ of $\nfeat=13$. 

For \vxDNN the $\vx_{i}$ are concatenated for all $i$ assuming two more jets in the NA class. For events with fewer than two jets in the NA class the foreseen features are initialized with zero. Together with the relational information between all potential objects in an event, this results in a dimension of \vxDNN of 175 features, of which up to 47 features might potentially be filled with zeros. 

In this configuration the information about the association to the partonic final state is presented in the form of one-hot encoding to both NN architectures. To confirm to what extent the DNN may infer this information already from the position of the $\vx_{i}$ within \vxDNN we also investigate configurations of the DNNs without this information in form of on-hot encoding, resulting in a reduced input vector \vxDNNred with dimension 105. All input features in use are listed in Table~\ref{tab:input_features}.
\begin{table}[t]
    \centering
    \caption{Input features used for the studies described in the text. Columns 2--4 indicate whether a given feature is continuous, discrete, or presented via one-hot encoding. The given features form the feature vectors $\vx_{i}$ per object $i$. For the DNNs the $\vx_{i}$ for potentially ten selected objects are concatenated into an extended feature vector \vxDNN, according to their association to the partonic final state; for events with fewer than ten objects the $\vx_{i}$ of the missing objects are filled with zeros. In addition, relational information between all potential objects is added to \vxDNN. In a reduced configuration, the one-hot encoded information about the association of the objects to the partonic final state is omitted to form a reduced input vector \vxDNNred to the DNNs.}
    \begin{tabular}{lcccl}
        \hline
        Input feature & Continuous & Discrete & One-hot & Comment \\
        \hline
        $m$ & $\checkmark$ & $-$ & $-$ & \hspace{-1em}\rdelim\}{6}{*}[Primary features] \\
        $E$ & $\checkmark$ & $-$ & $-$ &  \\
        $\pt$ & $\checkmark$ & $-$ & $-$ & \\
        $\phi$ & $\checkmark$ & $-$ & $-$ & \\
        $\eta$ & $\checkmark$ & $-$ & $-$ & \\
        $\beta$ & $-$ & $\checkmark$ & $-$ & \\
        LEP & $-$ & $-$ & $\checkmark$ & Assoc. to \Pell \\
        MET & $-$ & $-$ & $\checkmark$ & Assoc. to MET \\
        ADDB & $-$ & $-$ & $\checkmark$ & Assoc. to additional \Pb quarks \\
        HTB & $-$ & $-$ & $\checkmark$ & Assoc. to $\Pt_{\Pb_{\mathrm{had}}}$ \\
        LTB & $-$ & $-$ & $\checkmark$ & Assoc. to $\Pt_{\Pb_{\mathrm{lep}}}$ \\
        HTQ & $-$ & $-$ & $\checkmark$ & Assoc. to $\Pq_{\PW_{\mathrm{had}}}$ \\
        NA & $-$ & $-$ & $\checkmark$ & Not associated \\
        \hline
    \end{tabular}
    \label{tab:input_features}
\end{table}
All non-integer features are standardized to a mean of zero and a standard deviation of one.

\section{Results}
\label{sec:benchmark-tests}

To be able to draw fair conclusions from a comparison of different NN ar\-chi\-tec\-tures (of potentially different complexity) special care has to be taken for this comparison to be based on the same ground. For this study we have focused on a common choice of non-tunable (hyper-)parameters, which are not subject to the NN training, as well as on an equal level of information primarily passed to the NNs, through training conditions and input features. 

An inevitable difference remains in the organization and layering of hidden nodes, which when kept similar, may well lead to a different number of TPs and therefore a priori different expressiveness of the NN models. Vice versa, keeping the number of TPs similar, implies differences in the layering and organization of hidden nodes. Since differences of one or the other kind may not be overcome, both configurations, (i) similar layering of hidden nodes; and (ii) similar number of TPs, are studied. In any case, the enlarged size of \vxDNN (\vxDNNred) with respect to \vx will give higher emphasis to the first DNN layer compared to the corresponding GNN architecture. In addition, the DNN architecture features the less complex pre-processing of the inputs, since it does not imply the creation of graphs. On the other hand, the potentially more constrained GNN may have advantages over the DNN architecture in terms of convergence properties of the training.

\subsection{Comparison of neural networks with similar layering of hidden nodes}
\label{subsec:comp_architecture}

\subsubsection{Neural networks with one layer of hidden nodes}

\begin{figure}
    \centering
    \includegraphics[width=0.95\textwidth]{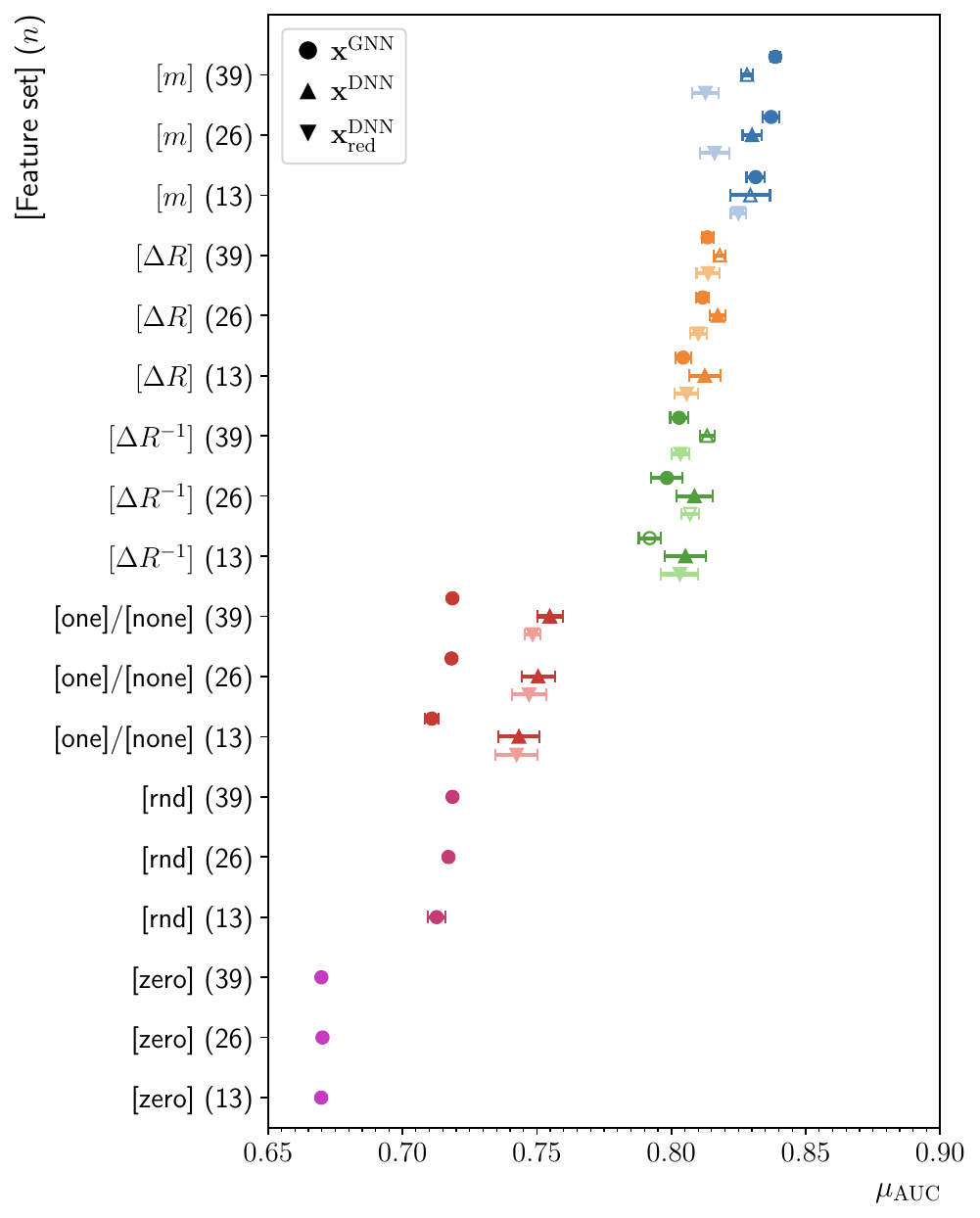}
    \caption{Mean ROC-AUC \muAUC as obtained for 18~different configurations of GNN and 24~corresponding configurations of DNN models with $\nconv=1$. The labels in brackets on the vertical axis indicate the use of relational information, as discussed in Section~\ref{sec:feature_space}, the numbers in parentheses correspond to the choices of \nemb. The circles refer to GNN and the upward (downward) pointing triangles to DNN models with a default (reduced) set of input features \vxDNN (\vxDNNred), as discussed in Section~\ref{sec:feature_space}. For better readability, markers of the same configuration are shifted vertically along the y-axis. The bars are obtained from the sample variance of an ensemble, as described in Section~\ref{subsec:training_setup}. Those NN architectures which belong to the same choices of varying parameters are spatially grouped and shown in the same color. Open markers indicate that significant outliers of the corresponding distribution of ROC-AUC values have been removed from the calculation of \muAUC and its variance, as described in the text.}
    \label{fig:results_comp_arch_1hl}
\end{figure}

A first comparison of GNN with corresponding DNN models, based on a similar layering of hidden nodes, is shown in Figure~\ref{fig:results_comp_arch_1hl}. 
The metric by which to judge the success of an NN to fulfill the task is chosen to be the mean of the ROC-AUC \muAUC based on the training setup, as described in Section~\ref{subsec:training_setup}. The results are presented for the variation of parameters as summarized in Table~\ref{tab:parameter_choices}. For the presentation in Figure~\ref{fig:results_comp_arch_1hl} a simple architecture with one hidden layer for the DNN models and one GraphConv operation of the GNN models ($\nconv=1$) has been chosen. For the GNN architecture this implies that there is only one exchange of information across adjacent nodes of $\mathcal{G}$, i.e., each node receives information only of its nearest and not the next to nearest neighbor in $\mathcal{G}$. Since $\mathcal{G}$ has been chosen to be fully connected, there is no strict suppression of information that way, in the sense that each node $i$ receives information from any other node $j\neq i$ in $\mathcal{G}$. The input features presented to each corresponding NN architecture are chosen, as described in Section~\ref{sec:feature_space} and summarized in Table~\ref{tab:input_features}. 

The results for the GNN models are represented by circles, the results for the DNN models with \vxDNN (\vxDNNred) as input vector by upward (downward) pointing triangles. The bars associated with the points indicate the uncertainty \DmuAUC in \muAUC due to random choices in the training, as described in Section~\ref{subsec:training_setup}. Open markers indicate that at least one training repetition in an ensemble has been removed as an outlier from the calculations of \muAUC and \DmuAUC. Candidates for outliers have been identified by their \muAUC values exceeding $1.5\,\DmuAUC$, as obtained from the full ensemble. An outlier candidate has been definitely removed and the ensemble size reduced accordingly, if doing so changed \DmuAUC by a value of at least 0.0025. Following this procedure, 54~outliers have been removed from a total of 1840, which corresponds to a rate of 2.9\%. Split by NN architectures, it corresponds to 17 (37) removed outliers for GNN (DNN) models from a total of 800 (1040). In no case more than two outliers have been removed from the original ten training repetitions of any individually model.

On the $x$-axis of the figure the corresponding values of \muAUC, ranging from 0.67 to 0.83, are shown. On the $y$-axis the individual NN models are labelled, such that brackets indicate, what relational information between final state objects has been used, and the values in parentheses indicate the choices of \nemb. 

We conclude that all training setups have succeeded in the sense that all NN models result in values of $\muAUC>0.5$. The worst separation of signal from background we observe for the GNN models for which no relational information is exploited, indicated by three groups of NN architectures shown in pink, purple and red colors in the lower part of the figure. It is noted that the feature set \zero refers to the case where the node convolution in the GraphConv operation is forcefully suppressed, and deliberately no information across nodes is exchanged at all. We anticipate that this approach counteracts the whole GNN idea. We still keep this configuration as part of the study, to gauge the effect and importance of the GraphConv operation itself. Compared to the feature set \zero, the feature sets \one and \rnd single out cases in which information exchange across nodes takes place, but no real relational information is associated with it. Instead, the embedding spaces of the individual nodes are just mixed without particular prevalence. We note that even when allowing node convolution the GNN architecture falls significantly behind the comparable DNN architecture, even with a reduced input vector \vxDNNred, as long as no mindful relational information across adjacent nodes $i$ and $j$ is provided according to $\omega_{ij}$. This is true for assigning (\one) the same or (\rnd) random weights to each edge, irrespective of the expressiveness of the NNs, indicated by \nemb. The superiority of the DNN architecture in this case cannot be attributed to the additional information about the jet parton association, since this information is available to all NN architectures under study. In particular it is passed on to the GNNs, in the explicit form of one-hot encoding, which is not even the case for the DNNs with \vxDNNred as input vector. At this occasion, we note that the additional (and in fact in this case redundant) information of the parton association in form of one-hot encoding to the DNN does not lead to a significant increase of \muAUC, compared to the implicit knowledge already provided by the positional information in \vxDNNred. Hence, if the way this information is presented to the DNN were of influence, this influence is not significant in the scope of our study. We also observe that \DmuAUC is considerably larger for the DNNs. 

In conclusion, if the advantage of the GNN  over the DNN architecture were that potentially excessive degrees of freedom in the DNN architecture are replaced by built-in constraints, the GNN architecture appears too confined, until these constraints are introduced mindfully. In turn the additional degrees of freedom of the DNN architecture result in a larger spread \DmuAUC due to random choices in the training setup.  

The upper part of Figure~\ref{fig:results_comp_arch_1hl} reveals that, as soon as a domain-knowledge motivated ranking of information exchange across neighboring nodes is introduced, the GNN architecture significantly gains in separation power. Also here, this gain comes with an increase in \DmuAUC. The choices of (green) \DRinv, (orange) \DR, and (blue) \Minv as weights leads to an increase in separation power in the given order, where for each choice the values of \muAUC can be grouped with a corresponding internal spread. On the other hand, a significant gain in \muAUC, when increasing the expressiveness of an NN within a certain configuration group in terms of \nemb, for the utilized NN models, is not observed.

We note that, as in the case of the positional encoding of parton association in \vxDNNred, all choices of relational information are intrinsic to the training sample and implicitly accessible to all NNs through their feature vectors. For \DR this is, e.g., the case through $\phi$ and $\eta$ in the primary features of each reconstructed object. However, the information of $\Delta\phi$ and $\Delta\eta$ between pairs of reconstructed objects appears too subtle in the high-dimensional feature space, so that none of the chosen architectures could grasp it without the assistance of an accordingly conditioned representation of \vx and \vxDNN, even from a training sample with more than 200,000 events. 

We further note that when turning the edge-weights of the GNN structure into TPs we did not obtain a separation of signal from background better than the domain-knowledge supported use of \Minv. At the same time we observed a significantly increased spread in the achieved separation power based on random choices of the training setup. 

Our physics prior assigns more physical meaning to the choice of \DR over \DRinv, since due to causality we expect a closer relation between objects with smaller than larger spatial distance in \DR. The observation that both choices of relational information lead to nearly similar results in \muAUC we explain by a special characteristic of NN-based classification tasks in the given setup. For the NN decision, downgrading information from further-away objects is equivalent to upgrading close-by objects.   
The fact that corresponding DNN architectures follow the trends of the GNN architectures, as long as equipped with the same information, supports the assumption that it is this additional relational information between objects, rather than the GNN-specific operation of mixing features across nodes that leads to the increase of \muAUC. 

We note that, consistently for all architectures, the highest values of \muAUC are obtained with an energy-weighted distance measure like \Minv, which again follows our prior physics intuition. It is noteworthy that in this case the DNN architecture with \vxDNNred seems to significantly lose in separation power, compared to the other architectures. In fact \muAUC even decreases for increasing values of \nemb. Also \DmuAUC appears consistently higher for all DNN compared to the GNN architectures. These observations may be interpreted as indications of the advantage of careful guidance of the NN training- and model-setup over just confronting a highly expressive NN architecture, represented by a large number of TPs, with an even excessively large training sample. This guidance may be provided through the choice and representation of input features, as well as through the choice of a more constrained NN architecture. We note that the highest value of \muAUC with the smallest spread \DmuAUC is indeed obtained from the GNN architecture with highest expressiveness, given for $\nemb=39$. 

\subsubsection{Neural networks with two layers of hidden nodes}

\begin{figure} [t]
    \centering   \includegraphics[width=0.51\textwidth]{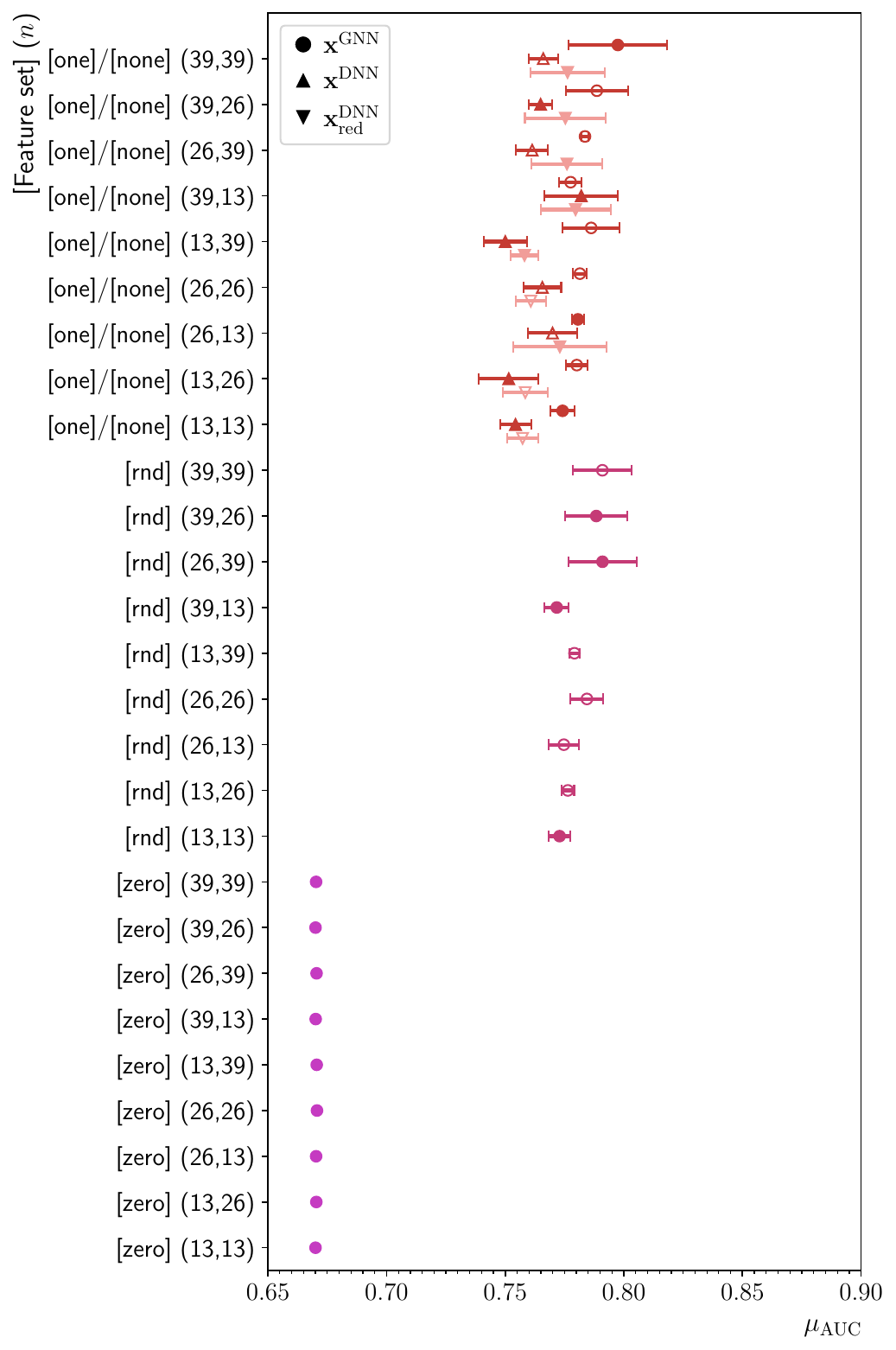}
    \includegraphics[width=0.48\textwidth]{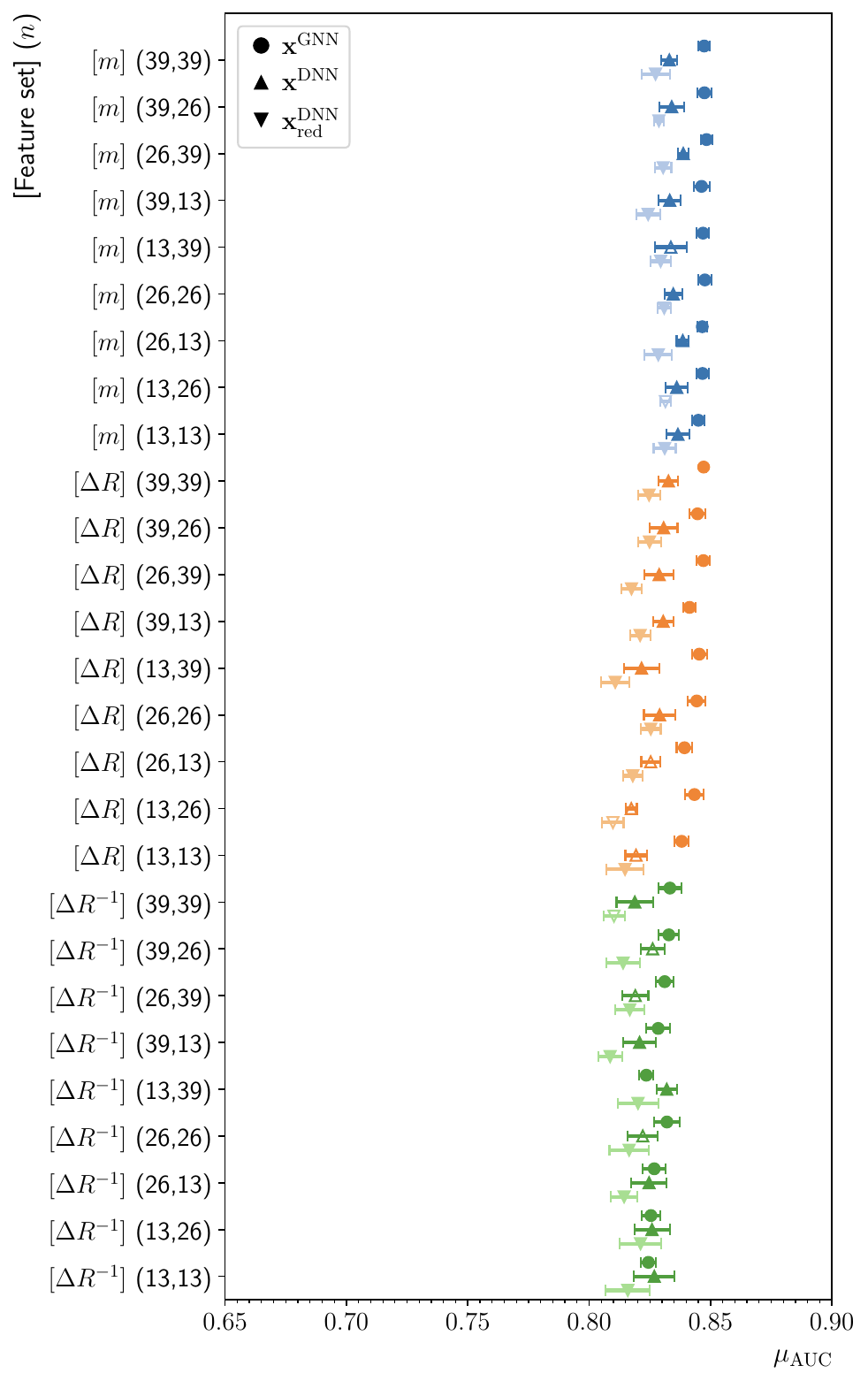}
    \caption{Mean ROC-AUC \muAUC as obtained for 54~different GNN and 72~corresponding DNN models with $\nconv=2$. The labels in brackets on the vertical axis indicate the use of relational information, as discussed in Section~\ref{sec:feature_space}, the numbers in parentheses correspond to \nemb. The circles refer to GNN and the upward (downward) pointing triangles to DNN models with a default (reduced) set of input features \vxDNN (\vxDNNred), as discussed in Section~\ref{sec:feature_space}. For better readability, markers of the same configuration are shifted vertically along the y-axis. The bars are obtained from the sample variance of an ensemble, as described in Section~\ref{subsec:training_setup}. NN architectures which belong to the same choices of varying parameters are spatially grouped and shown in the same color. Open markers indicate that significant outliers of the corresponding distribution of ROC-AUC values have been removed from the calculation of \muAUC and its variance, as described in the text.}
    \label{fig:results_comp_arch_2hl}
\end{figure}

Moving on to an NN architecture with $\nconv=2$ introduces the ambiguity of how to choose \nemb for each individual layer. To prevent any kinds of potential selection biases, all ways of allocating the tested values of $\nemb=13,\,26,\,39$ to the individual layers/embedding spaces are shown in Figure~\ref{fig:results_comp_arch_2hl}.

For the case of completely suppressed relational information for the GNN architectures (\zero) \muAUC remains lowest and unaffected by the choice of \nconv, as expected for a setup in which any information transfer through a GaphConv operation is deliberately suppressed. 
At the same time we generally observe that no superior choice of allocating \nemb across layers can be pointed to, throughout all tested architecture configurations. Especially the gain of using a configuration with $\nemb=(39,\,39)$ for both hidden layers/GraphConv operations over a configuration with $\nemb=(13,\,13)$ within a given architecture appears marginal. 

However, the increase in \nconv consistently mitigates the previously observed, clearly inferior separation of signal from background, of the GNN compared to the corresponding DNN architectures, for randomly (\rnd) and unweighted (\one) relational information. The \muAUC values of these groups of GNN models start to clearly supersede the \muAUC values of the one-layered DNN models with unweighted relational information, labeled by \none in Figure~\ref{fig:results_comp_arch_1hl}, even slightly taking the lead over the two-layered DNNs of the same kind, in terms of \muAUC. 

We note that all two-layered NN architectures without use of relational information still result in lower values of \muAUC than all tested one-layered NN models that profit from the use of relational information, as presented in Figure~\ref{fig:results_comp_arch_1hl}. At the same time, they are subject to an increased spread \DmuAUC compared to their one-layered counterparts, in most cases. In this sense, the wise choice of relational information outweighs the presumable advantage in expressiveness provided by $\nconv=2$, irrespective of the allocated values of \nemb. 

For the NN architectures including relational information, we observe no further, dramatic gains, with respect to their one-layered counterparts in \muAUC, apart from a slight advantage of the GNN over the corresponding DNN architectures that seems to become more manifest. While this advantage is below the 1\%-level it is still significant compared to \DmuAUC. 
\begin{figure}
    \centering    
    \includegraphics[width=0.8\textwidth]{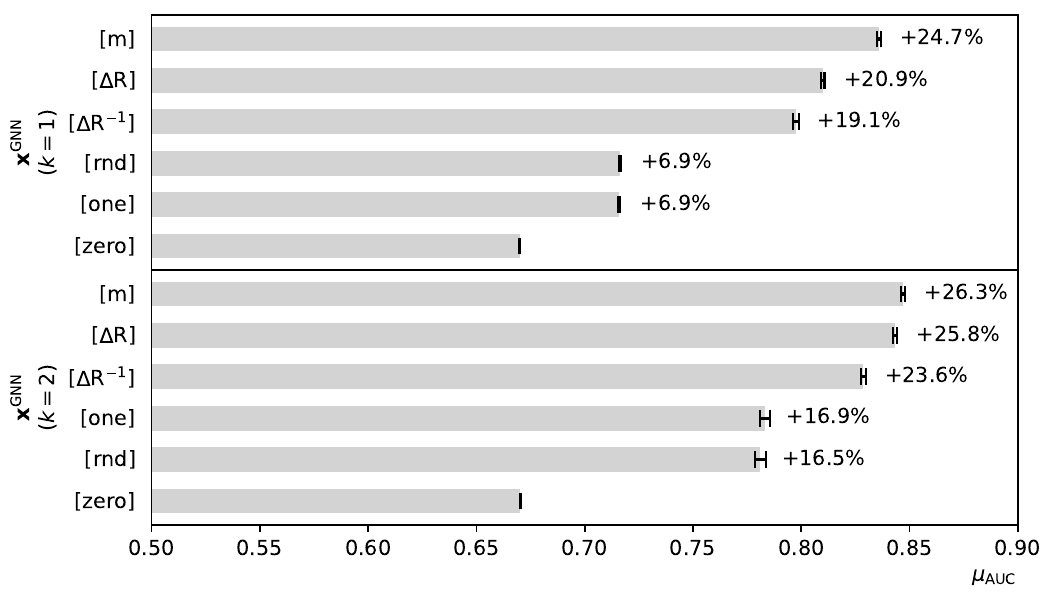}
    \caption{
        Summary of the achieved values of \muAUC for the GNN models with (upper half) one and (lower half) two GraphConv operations, with different use of relational information. For this summary, the associations of \nemb with the highest values of \muAUC in each group of GNN models have been used. The value of \muAUC is displayed on the $x$-axis. Improvements relative to the least separating GNN with no relational information at all (\zero) is given in numbers to the right of the bars. The use of relational information, as defined in Table~\ref{tab:parameter_choices}, is indicated in brackets, on the $y$-axis.
    }    \label{fig:results_comp_arch_1hl_edge_weights_GNN}
\end{figure}
\begin{figure}
    \centering    
    \includegraphics[width=0.8\textwidth]{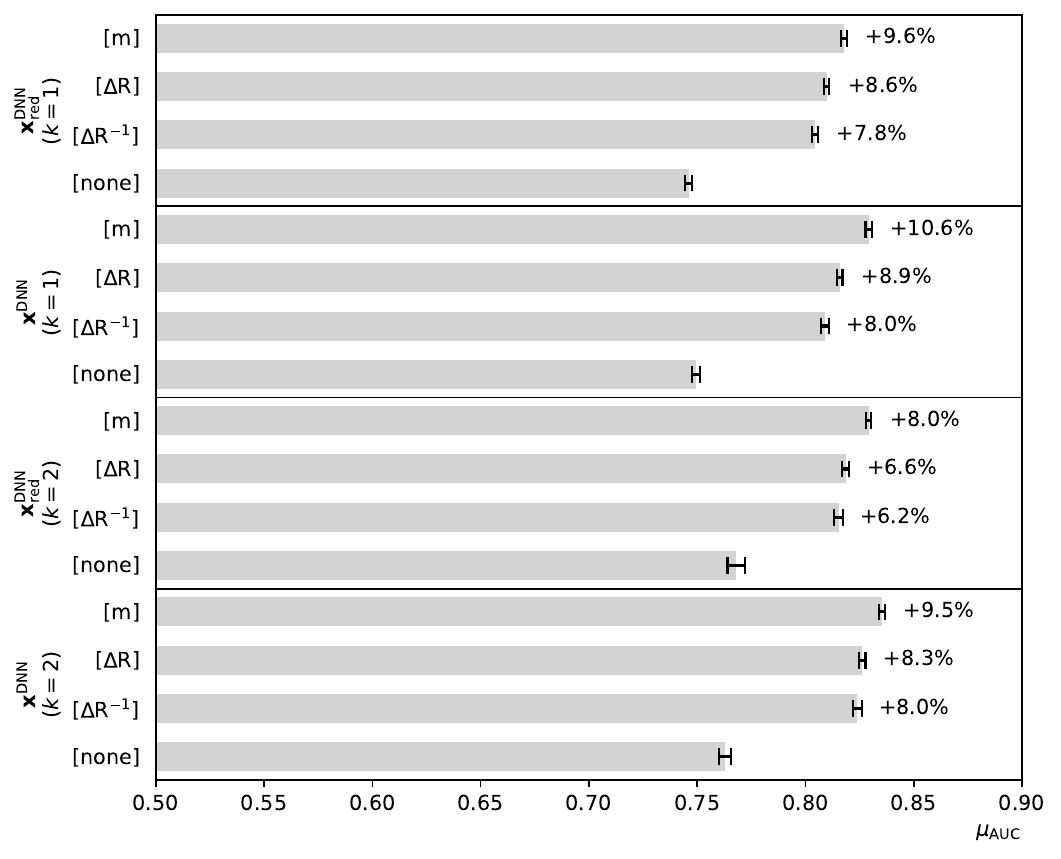}
    \caption{
        Summary of the achieved values of \muAUC for the DNN models with (upper half) one and (lower half) two hidden layers, with different use of relational information. For this summary, the associations of \nemb with the highest values of \muAUC in each group of DNN models have been used. The DNN configurations with \vxDNN and \vxDNNred are shown separately. The values of \muAUC are displayed on the $x$-axis. Improvements relative to the least separating DNN with no relational information (\none) is given in numbers to the right of the bars. The use of relational information, as defined in Table~\ref{tab:parameter_choices}, is indicated in brackets, on the $y$-axis.
    }    \label{fig:results_comp_arch_1hl_edge_weights_DNN}
\end{figure}
A summary of the achieved values of \muAUC for the one- and two-layered GNN models is shown in Figure~\ref{fig:results_comp_arch_1hl_edge_weights_GNN}. An equivalent summary for the corresponding DNN models is shown in Figure~\ref{fig:results_comp_arch_1hl_edge_weights_DNN}.

From the study we conclude that the external information of the $\omega_{ij}$ seems to give slight advantages to the GNNs with $\nconv=2$. We note that two subsequent GraphConv operations indeed convey more information than a DNN model with two hidden layers. Viewing \DRinv, \DR, and \Minv as distance measures, the first GraphConv operation conveys information about the nearest neighborhood of each $i$. The second GraphConv operation conveys information about the nearest neighborhood of the nearest neighbors, which is not the same as in the case of the first operation. This information is indeed not primarily accessible to the DNN architectures, but it emerges from the definition of the GraphConv operation. Along this line, once again, we conclude that not the mixing of features across neighboring nodes $i$ during the GraphConv operation, but the additional (implicit) information accessible to the GNN through this operation is the source for the slight gain in \muAUC. This interpretation is supported by the observation that an increase in \nemb does not lead to any significant improvements in \muAUC despite the increase in expressiveness of the models. This also explains why the GNN models with constant (\one) or randomly associated (\rnd) weights suffer in their performance: Through both ways of assigning weights across the nodes $i$ the information about any kind of distance measure across nodes, in what ever space, is omitted. 

A summary of the GNN and DNN configurations with the highest values of \muAUC is given in Table~\ref{tab:NN-setup-summary}.

\begin{table}[b]
\centering
\caption{
    Summary of \nemb, number of TPs (\NTP), and \muAUC of the (upper part) GNN and (lower part) DNN models with the highest results in \muAUC. For the DNN models a number of effective TPs \NTPeff, as defined in the text, is also given in parentheses. Corresponding summaries are given for the cases of $\nconv=1$ and 2. In all cases \Minv has been used as relational information between nodes/physics objects. Also shown are the configurations of three additional NN models discussed in Section~\ref{subsec:comparable_number_of_TPs}: the \expGNN model with \NTP comparable within 1\% with $\NTP(\DNNtwoL)$ and the \restrDNN (\restrDNNeff) model with \NTP (\NTPeff) comparable within 1\% with $\NTP(\GNNtwoL)$.
    }
{\renewcommand{\arraystretch}{1.4} 
\begin{tabular}{llcr@{$\pm$}rl}
\hline
NN arch. & \multicolumn{1}{c}{\nemb} & \NTP(\NTPeff) & \multicolumn{2}{c}{\muAUC} & Label \\
\hline
\hspace{+0.5em}\ldelim\{{3}{*}[GNN$\hspace{0.5cm}$] 
& (39)             & 1093 & 
\num[round-mode=places, round-precision=4]{0.8387019565424094} &
\num[round-mode=places, round-precision=4]{0.0005987504259} 
& \GNNoneL \\
& (26, 39)         & 2809 & 
\num[round-mode=places, round-precision=4]{0.8483569369144194} &
\num[round-mode=places, round-precision=4]{0.0007726407041} 
& \GNNtwoL \\
& (29, 28, 29, 29) & 5829 & 
\num[round-mode=places, round-precision=4]{0.854602560655694 } & 
\num[round-mode=places, round-precision=4]{0.0005777567118} & \expGNN \\
\hspace{+0.5em}\ldelim\{{4}{*}[DNN$\hspace{0.5cm}$] 
& (26)             & 5773 (4753) & 
\num[round-mode=places, round-precision=4]{0.8300379444445589} &
\num[round-mode=places, round-precision=4]{0.001116829909}  
& \DNNoneL \\
& (26, 39)         & 6839 (5819) & 
\num[round-mode=places, round-precision=4]{0.8387656640349572} &
\num[round-mode=places, round-precision=4]{0.0007282520914}   
& \DNNtwoL \\
& (13, 14, 14)     & 3294 (2784) & 
\num[round-mode=places, round-precision=4]{0.8400395495410187} & 
\num[round-mode=places, round-precision=4]{0.0005809342368}  
& \restrDNNeff \\
& (11, 10, 11, 11) & 2816 (2385) & 
\num[round-mode=places, round-precision=4]{0.8386413715083743} &
\num[round-mode=places, round-precision=4]{0.0007411985891}  
& \restrDNN \\
\hline
\end{tabular}
}
\label{tab:NN-setup-summary}
\end{table}

\subsection{Neural networks with comparable numbers of trainable parameters}
\label{subsec:comparable_number_of_TPs}

As stated before, two principally different NN architectures lack comparability in the sense that it might be more natural to pick up certain information from the training sample through one or the other architecture. As long as it addresses an intrinsic property of one or the other architecture, such a difference is part of the benchmark comparison. If, on the other hand, such an inequality results from withholding primary information from one or the other architecture, or potentially inappropriate advantages in terms of expressiveness, an effort should be made to study and estimate the effect of it. 

We have identified and noted such an inequality, in the beginning of Section~\ref{sec:benchmark-tests}, in terms of the number of TPs (\NTP), which turns out to be naturally higher for the DNN compared to the GNN architecture, due to the usually much larger input layer. We therefore complete our study by three additional setups, for which we drop the restrictions on the layered structure of the trained NN models in favor of comparable numbers of TPs. 

As shown in Table~\ref{tab:NN-setup-summary}, the GNN model with the highest value of \muAUC (labelled as \GNNtwoL) is based on $\nconv=2$ and $\nemb=(26,\,39)$, with $\NTPGNN=2809$ and a value of $\muAUC=0.8484\pm0.0008$. The DNN with the highest result of \muAUC (labelled as \DNNtwoL) is based on the same configuration in terms of \nconv and \nemb, with $\NTPDNN=6839$ and a value of $\muAUC=0.8388\pm0.0007$. One may argue that for the DNN model, not the full set of TPs is really actively contributing to the solution of the task, since part of the input space is regularly filled with zeros, e.g., if fewer than eight jets are selected in an event. Therefore, we estimate, in addition to \NTP, a number of effective TPs (\NTPeff) from the product of \NTP with the average number of nonzero input nodes in \vxDNN, evaluated on the training dataset. This results in a value of $\NTPeffDNN=5819$. 

In a first approach we survey varying DNN structures with \NTP (\NTPeff) comparable to \NTPGNN. We do this based on the following algorithm: We allow $\nconv\leq4$ and any number of nodes per hidden layer (\nemb). Of all DNN configurations for which \NTPGNN is matched by \NTP (\NTPeff) within a margin of 1\%, the model with the smallest spread of \nemb across layers is selected. If no DNN configuration with \NTP (\NTPeff) within a 1\% margin of the target value can be found the closest possible model is chosen. This situation occurs only in models with one hidden layer.

This procedure ensures a homogeneous structure of hidden layers. As a result, e.g., a DNN configuration with $\nconv=4$ and $\nemb=(11,\,10,\,11,\,11)$, with $\NTP=2816$, further on referred to as \restrDNN, is preferred over a model with $\nconv=4$ and $\nemb=(10,\,8,\,14,\,24)$, even though the latter results in an exact match with \NTPGNN. The \restrDNN model achieves a value of $\muAUC=0.8386\pm0.0007$. A second DNN configuration with $\nconv=3$ and $\nemb=(13,\,14,\,14)$, with $\NTPeff=2784$ within the 1\% margin of \NTPGNN is also considered and further on referred to as \restrDNNeff. This model achieves a value of $\muAUC=0.8400\pm0.0006$. We observe that, although the \restrDNN (\restrDNNeff) model uses only 41\% (48\%) of \NTP (\NTPeff) of \DNNtwoL, this does not result in any significant loss in separation power, after training.

In a second approach we survey varying GNN structures with \NTP comparable to \NTPeffDNN. For this purpose we exploit the same algorithm as described above, resulting in a GNN with $\nconv=4$ and $\nemb=(29,\,28,\,29,\,29)$ with $\NTP=5829$, further on referred to as \expGNN. This model achieves a value of $\muAUC=0.8546\pm0.0006$. It reveals the highest value of \muAUC across all tested models. The difference in \muAUC with respect to other NN configurations is only at the 1\%-level, but it is still significant in terms of \DmuAUC. 

From this finding we conclude that the GNN model with the same expressiveness as a maximally comparable DNN must have intrinsic advantages over the DNN model in extracting additional information from the given, large training sample. For the benchmark setup in use, this advantage is small but significant in the scope of the study. It emerges after external augmentation with an energy-weighted distance measure like \Minv between the input objects/nodes, and more clearly manifests itself in the study for $\nconv>1$. We anticipate that this gain originates from the hierarchically structured information about nearest neighbors and the nearest neighborhood of nearest neighbors of node $i$, when viewing the edge weights $\omega_{ij}$ as a distance measure. This information is an intrinsic property of the GNN model and not easily accessible through the more simplistic DNN structure. An increase of TPs of the simpler DNN structure does not compensate for this informational advantage. In this interpretation the gain of \expGNN over all other configurations should mostly be attributed to the increase in \nconv over the association of \nemb per GraphConv operation. The parameter choices of the \restrDNN, \restrDNNeff, and \expGNN models and correspondingly achieved values of \muAUC are also given in Table~\ref{tab:NN-setup-summary}.

\subsection{Convergence behavior}

\begin{figure}[t]
    \centering
    \includegraphics[width=0.49\textwidth] {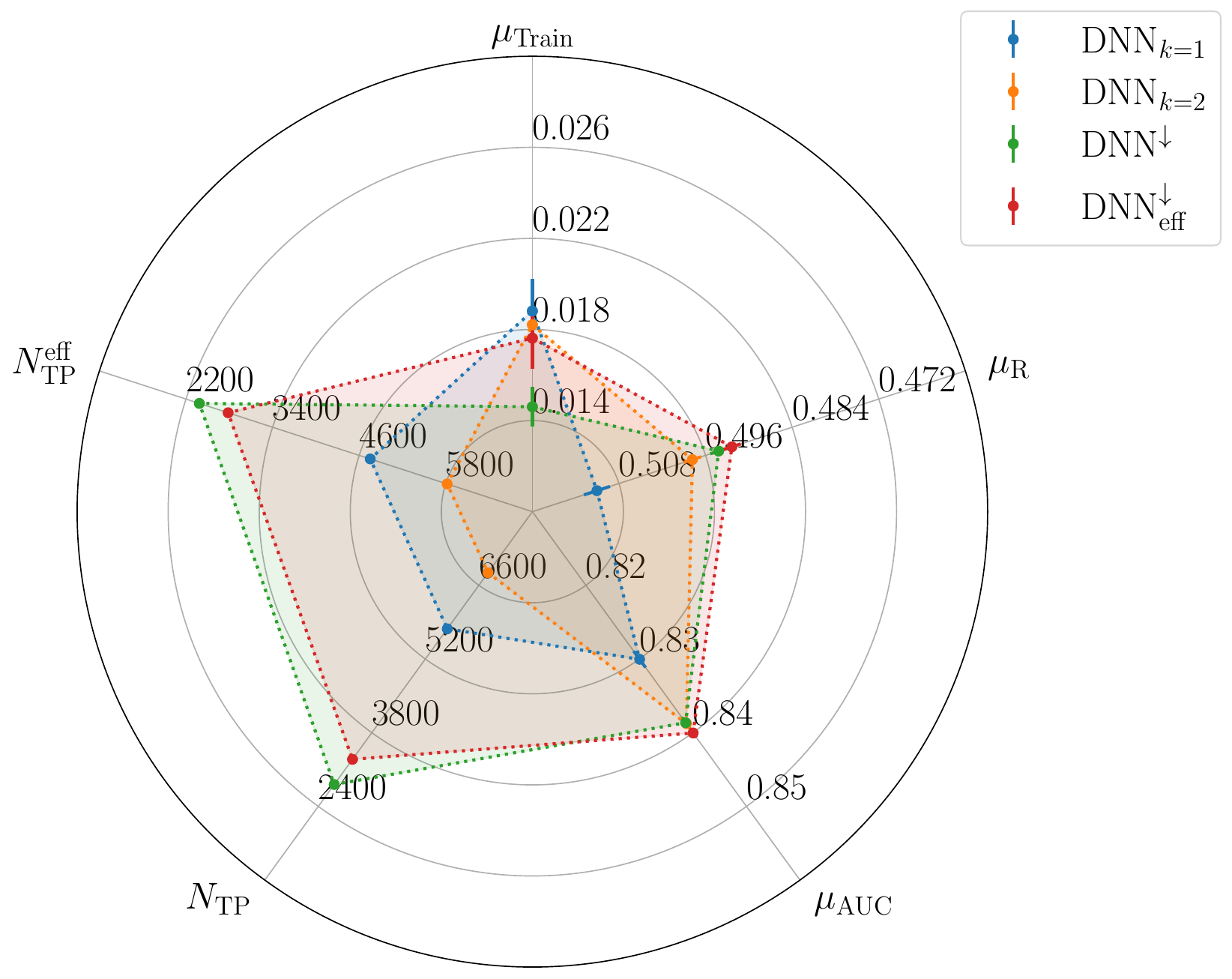}
    \includegraphics[width=0.49\textwidth] {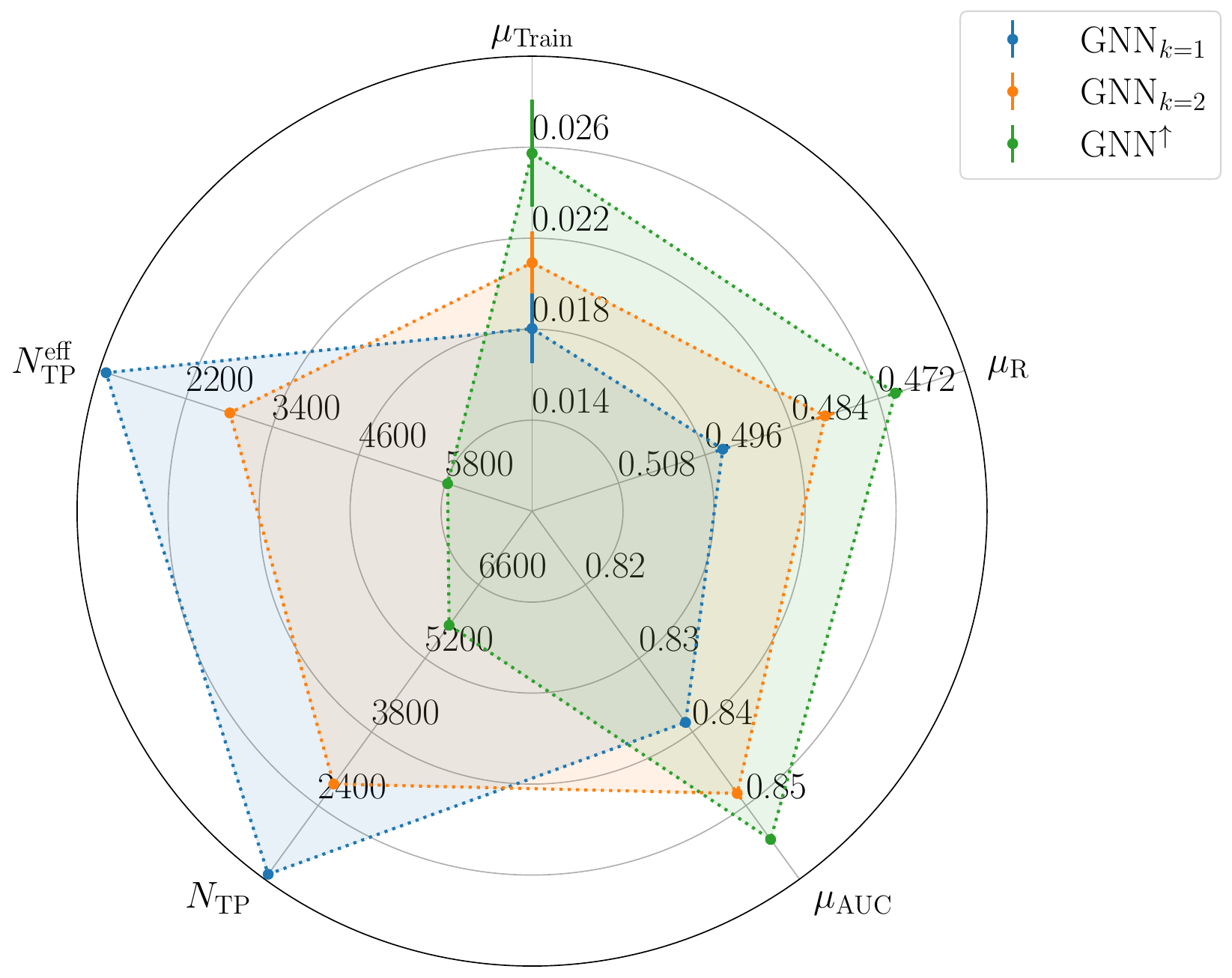}
    \caption{
        Visualization of GNN and DNN models with the highest values in \muAUC and  $\nconv=1$ (\GNNoneL, \DNNoneL) and $\nconv=2$ (\GNNtwoL, \DNNtwoL). Also shown are the DNN models \restrDNN and \restrDNNeff with a comparable number of (effective) TPs as the \GNNtwoL model, and the GNN model \expGNN with a comparable number of effective TPs as the \DNNtwoL model. Five different metrics to evaluate the properties of an NN model are shown: \NTP (\NTPeff), the mean convergence rate \mutrain, the mean empirical risk evaluated on the test sample \muR, and \muAUC. The spanned area in the figures indicates the capability of an NN model to fulfill the task.
        }
    \label{fig:spider_plots}
\end{figure}

In this study we have investigated the capacities of GNN and DNN models to fulfill their primary target, i.e., to provide the best possible solutions to the classification task defined in Section~\ref{subsec:task_definition}. We anticipate that, in particular in practical life, the properties of an NN architecture may be evaluated in other terms, viz. the mean of the training speed \mutrain, which we evaluate as the inverse of the epoch with the highest value of the ROC-AUC on the validation dataset and the mean of the empirical risk obtained from the test dataset \muR, which we take as a measure of the generalization property of the given NN model under study. To conclude our studies we provide a visualization of these properties and all other properties of the NN models that have been discussed throughout the paper so far, in Figure~\ref{fig:spider_plots}. 
In Figure~\ref{fig:spider_plots} (left) \mutrain, \muR, \muAUC, \NTP, and \NTPeff for the \DNNoneL, \DNNtwoL, \restrDNN, and \restrDNNeff models are shown, on five independent axes. The axes are defined such that values closer to the common origin of the figure are disfavored. This is in particular true for \NTP, \NTPeff, and \muR, where the values are given in descending order when moving away from the origin. In this sense a larger size of the correspondingly colored area indicates the larger capability of a given NN model to adequately solve the presented task. In Figure~\ref{fig:spider_plots} (right) the same quantities are shown for the \GNNoneL, \GNNtwoL, and \expGNN models. The axes ranges are kept the same to ease comparison between both architectures. 

\begin{figure}[t]
    \centering
    \includegraphics[width=0.99\textwidth] {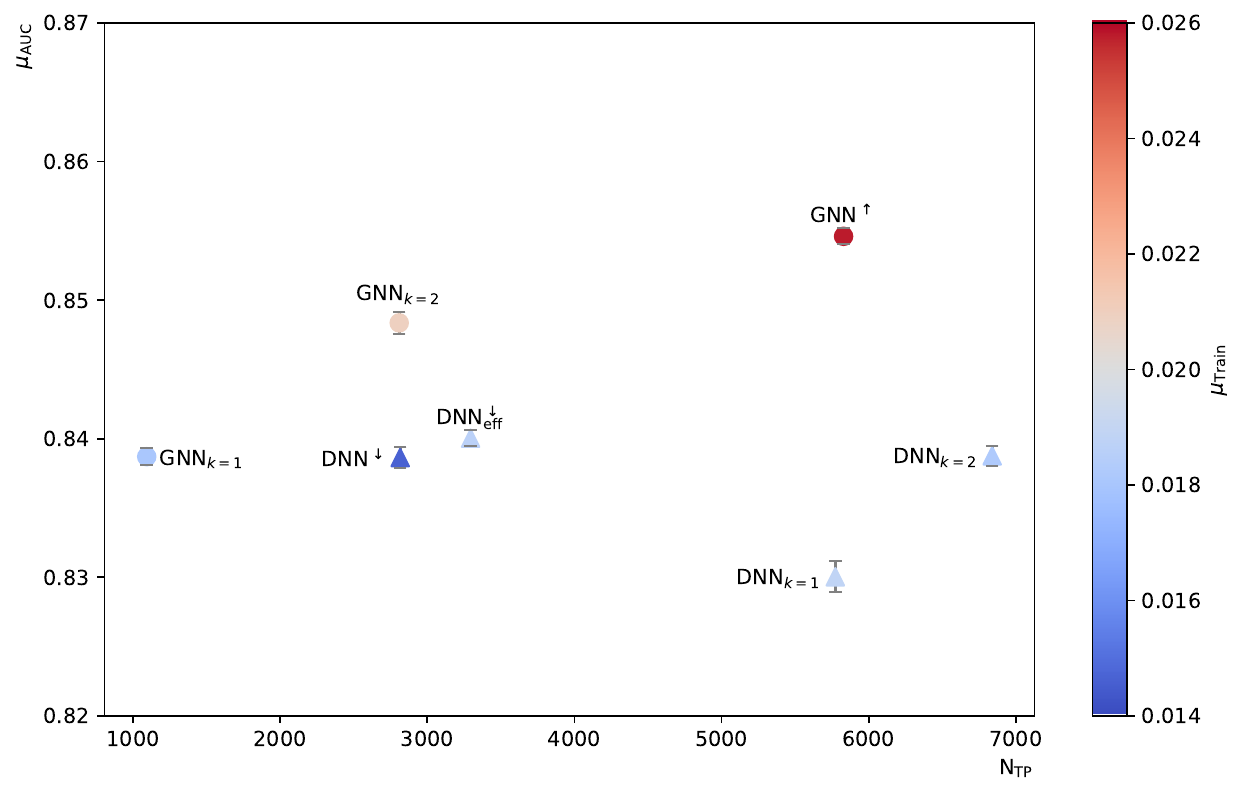}
    \caption{
        Overview of the GNN and DNN models with $\nconv=1$ and 2, and the highest achieved values of \muAUC, as well as a GNN model with an increased number of TPs (\expGNN) and a DNN model \restrDNN (\restrDNNeff) with a restricted (effective) number of TPs. The GNN models are indicated by circles and the DNN models by triangles. The models are shown in a three-dimensional space built from \muAUC, \NTP and \mutrain. The bars in \muAUC are obtained from the sample variance of a training ensemble, as described in Section~\ref{subsec:training_setup}. The quantity \mutrain, indicated by the color code of the points corresponds to the inverse of the epoch with the highest value of the ROC-AUC on the validation dataset.}
    \label{fig:plot_3d_money}
\end{figure}

In terms of \mutrain the DNN models usually fall behind their corresponding GNN counterparts. This finding, as well as the observation that the GNN models usually achieve a comparable or slightly larger value of \muAUC, after training, indicates the assumed effect of guiding the convergence by constraints, which are built-in to the GNN architecture. This property allows the GNN architecture not only to converge to a solution that is equally good or slightly better than the solution found by corresponding DNN architectures, but also to converge faster and typically with a smaller number of TPs. We note though that a clear correlation between \mutrain and \NTP (\NTPeff) cannot be deduced from our study. We understand this situation such that a less expressive NN model, with fewer degrees of freedom for the minimization process, may well lead to a more pronounced landscape of the expected risk and thus reduced \mutrain. In addition we note that also \muR for the DNN models falls behind, compared to their GNN counterparts. Here we observe the benefit of a regularizing effect of the built-in constraints, which correlates with reduced numbers of (effective) TPs. It is obvious that an NN model with more TPs reveals a higher vulnerability to specific properties of the training sample. A summary of all NN configurations that have been discussed in this section is shown in Figure~\ref{fig:plot_3d_money}.

\section{Summary}
\label{sec:summary}

With this paper we have made an effort to put the comparison of graph neural network (GNN) and equivalent fully-connected feed-forward neural network (DNN) architectures on a maximally fair ground. Under the laboratory conditions of a high-energy physics process of interest, at the CERN LHC, we have controlled the definition of the task, choice of non-tunable (hyper-)parameters of the models, which are not subject to the training, and amount of information primarily presented to the neural network models, through their input feature vectors. 

Within the scope of the study we have demonstrated clear evidence that the presumable advantage of the more complex GNN over an equivalent DNN structure does not originate from an uncontrolled mixture of features in the embedding space of the graph nodes, but from the extra relational information between nodes, that we have added based on domain-knowledge. Without this extra knowledge the GNN models fall behind equivalent DNN models in terms of their capability to separate signal from background. Neither are the GNN models superior in terms of their separation power to equivalent DNN models, as long as these are equipped with the same information in the input space. 

Both, the built-in permutation invariance and the circumstance that the GNN architecture a priori is not bound to a fixed number of nodes might be viewed as advantages. They might also have positive influence on the convergence behavior of the training. On the other hand it cannot be deduced that either of these properties significantly contributes to an increase, e.g., in the power to separate a given signal from background. Any advantage of the GNN over the DNN architecture that we observed in our studies could be traced back to the access to more information, which, when given to the other architecture lead to the same performance also for the other architecture.   

The real advantage of the GNN over the DNN structure emerges as soon as more than one GraphConv operations are applied, during which the GNN structure naturally accesses more relational information between nodes than a DNN has access to. In the configurations investigated in our study this gain is tied to the use of relational information that can be interpreted as a distance measure to define proximity between two nodes. Apart from that we observe significant advantages of the GNN over the DNN architecture in terms of convergence and generalizability that we attribute to a level of built-in implicit constraints to the GNN model resulting in a better ratio of accessible information over trainable parameters of the model. We anticipate that these advantages might be more pronounced the more hierarchical the training data are. We assume that this property is the basis for the success of GNN structures when applied to particle physics jets, which are highly hierarchical objects. In conclusion we expect the highest gain of a GNN over a DNN structure for tasks based on hierarchically structured data, ideally based a known distance measure. 

\section*{ACKNOWLEDGEMENTS}
This research was supported by the German Federal Ministry of Education and Research (BMBF) under grant 05H21VKCCB.


\bibliographystyle{unsrt}
\bibliography{references}   

\end{document}